\documentstyle[preprint,revtex]{aps}	
\begin{document}
%
\def\eslt{E\llap/_T}
\def\to{\rightarrow}
\def\te{\tilde e}
\def\tg{\tilde g}
\def\tga{\tilde \gamma}
\def\tnu{\tilde\nu}
\def\tq{\tilde q}
\def\tt{\tilde t}
\def\tm{\tilde m}
\def\tb{\tilde b}
\def\tw{\tilde W}
\def\tz{\tilde Z}
\def\tf{\tilde f}
\def\tl{\tilde l}
\def\tu{\tilde u}
\def\td{\tilde d}
%
%
\preprint{FSU-HEP-930527}
\preprint{UH-511-767-93}
\preprint{\today}
\begin{title}
NEW SIGNALS FOR GLUINOS AND SQUARKS\\ OF SUPERGRAVITY\\
AT THE TEVATRON COLLIDER
\end{title}
\author{Howard Baer$^1$, Chung Kao$^1$ and Xerxes Tata$^2$}
\begin{instit}
$^1$Department of Physics,
Florida State University,
Tallahassee, Florida 32306 USA
\end{instit}
\begin{instit}
$^2$Department of Physics and Astronomy,
University of Hawaii,
Honolulu, HI 96822 USA
\end{instit}
\begin{abstract}
Within the supergravity framework, sleptons are expected to be considerably
lighter than squarks if gluinos and squarks are approximately degenerate.
This can lead to a significant enhancement in leptonic branching ratio for
neutralinos, and sometimes, charginos.
Using ISAJET 7.0/ISASUSY 1.0, we evaluate the multilepton
signals from cascade decays of gluinos and squarks
produced at the Fermilab Tevatron $p\bar p$
collider assuming squark and slepton masses are related as in supergravity
models. We find observable cross sections for
spectacular event topologies: $m$-lepton + $n$-jet + $\eslt$ events
($m \leq 4$, $n \geq 2$) and same-sign dilepton + $\eslt$ events and
show that the SM backgrounds to the isolated trilepton, four
lepton and same-sign dilepton signals are very small. These signals can extend
the mass reach of the Tevatron experiments to $m_{\tg}\sim 300$~GeV if
$m_{\tg}\sim m_{\tq}$ and ${\cal O}(1000)$ pb$^{-1}$ of data are collected.

\end{abstract}
\pacs{PACS numbers: 14.80.Ly, 11.30.Pb, 13.85.Qk}
\newpage
%
%
\begin{narrowtext}

Supersymmetry (SUSY) provides a natural framework for the
incorporation of elementary scalars into theories of fundamental
processes, provided only that
sparticles are not much heavier than the electroweak scale. This has
provided ample
motivation to seriously consider\cite{REV} the implications of SUSY
for high energy
collider experiments. Further motivation for low energy SUSY is provided by the
observation\cite{UNIF} that the predictions of the simplest SUSY SU(5) model,
unlike those of its non-supersymmetric counterpart, are consistent with both
the measurements of the gauge couplings at LEP as well as with the experimental
constraints on the lifetime of the proton. In addition, SUSY theories
potentially provide a dark matter candidate if R-parity is conserved.

The systematic exploration of the experimental consequences of supersymmetry
necessitate appealing to a model. Almost always, phenomenological analyses are
performed within the framework of the Minimal Supersymmetric Model (MSSM) which
is the simplest supersymmetrization of the Standard Model (SM) in that it
contains the smallest number of new particles and new interactions necessary
for phenomenology. Within this framework, our ignorance of the mechanism of
SUSY breaking is parameterized by a number of soft SUSY breaking masses and
interactions. Without further assumptions\cite{ERICE} about the symmetries of
SUSY breaking interactions, this leads to a proliferation of model parameters
making phenomenological analyses intractable. Especially appealing are locally
supersymmetric theories where gravitational effects act as messengers of SUSY
breaking (assumed to occur in a different sector of the theory) to the
observable sector\cite{REV}. Since gravitational interactions are universal,
this leads to inter-relations amongst the SUSY breaking parameters at the
unification scale; for instance, a common soft-SUSY breaking scalar mass may be
sufficient  regardless of any assumptions about the gauge group. If we further
assume that there is no other physics between the unification and weak scales,
it is straightforward to evolve the SUSY breaking parameters down from the
unification scale to the low energy scale using renormalization group
methods\cite{RENORM} in order to sum the large logarithms arising from the
disparity of the two scales. It is worth remarking that the assumption of a
common scalar mass at the unification scale leads to an approximate degeneracy
of the squarks of the first two generations as required\cite{FLAVOUR} by the
phenomenology of K mesons. What makes supergravity models especially attractive
is that the logarithmic corrections also induce the breakdown of electroweak
symmetry, so that gravity serves to link the electroweak and the low energy
SUSY breaking scales.

In such scenarios, the
masses and couplings of all the sparticles can be computed\cite{SPECTRA} in
terms of just three additional parameters so that sparticle masses, and
hence their signals, are strongly correlated.
We focus our attention
only on squark and gluino signals at the Tevatron, assuming that sfermion
and gaugino masses are related as in supergravity\cite{REV}, but treating
other parameters
as in the MSSM. Thus, the predictions of supergravity  would be a special
case of our analysis. One of our main motivations for re-assessing these
signals derives from the observation that sleptons are significantly
lighter than squarks if squarks and gluinos have roughly comparable masses.
In this case, the leptonic decays of the neutralinos and charginos can
be significantly enhanced\cite{BT,LOPEZ} leading to an increase in the
leptonic signals resulting from the cascade decays\cite{CAS,BBKT} of
squarks and gluinos produced at the Tevatron. These signals, which have mainly
been studied for hadron supercolliders\cite{BBKT,SUPER}, could lead to novel
signatures even at the Tevatron\cite{BTW,BGH,SNOW}. With the CDF and D0
experiments
well underway, and with the possibility of the main injector upgrade in the
near future, we felt that a re-evaluation of these novel signatures, which
could
have an impact on the search for SUSY at the Tevatron, was warranted.

Towards this end, we have, together with Paige and Protopopescu, recently
incorporated\cite{BPPT} sparticle
production and decays as given by the MSSM into ISAJET\cite{ISAJET}
(version 7.0) which is a Monte Carlo
program for simulating high energy $pp$ and $p\bar p$ collisions.
For present purposes, it is sufficient to know that ISAJET
simulates the production of the gluino and five squark flavours (the t-squark
is not yet included because its decay patterns, and hence signatures, depend
on model-dependent t-squark mixing\cite{STOP}) with cross sections determined
in terms of $m_{\tq}$ and $m_{\tg}$ by QCD. These squarks and gluinos
then decay into jet(s) and any chargino or neutralino that is
kinematically accessible, with branching
fractions determined by the MSSM parameters ($m_{\tg}$, $m_{\tq}$,
$\mu$, $\tan\beta$, $m_{H_p}$) defined in Ref.\cite{BPPT} together with
$m_t$. The daughter charginos or neutralinos further decay until the cascade
terminates
in the lightest SUSY particle (assumed to be the lightest
neutralino, $\tz_{1}$) which escapes detection in the experimental
apparatus. We assume that the slepton masses
are determined by $m_{\tq}$ as, for example, in Ref.\cite{BT} and also
that the gaugino masses unify at an ultra-high scale\cite{REV}.
The leptonic decays, $\tw_i \rightarrow l\nu\tz_{1}$ and
$\tz_i \rightarrow l\bar{l}\tz_{1}$, of the charginos and
neutralinos produced at the intermediate stage of the cascades lead to the
production of isolated hard leptons in squark and gluino events resulting in
topologies with $n$-jets + $m$-leptons +$\eslt$. Initial and final state
parton showers, fragmentation and decays of heavy flavours, hadronization
effects, and finally, underlying event structure are modelled in ISAJET.

The prospects for detecting these leptonic signals at the Tevatron forms the
main focus of this study. Especially interesting are the trilepton\cite{BTW}
and same sign dilepton\cite{BGH,BTW,SNOW} signals, since SM backgrounds to
these are expected to be very small. We have improved on earlier studies of
these signals which were carried out using parton level Monte Carlo programs in
two significant respects. First, unlike the parton-level simulations, the
simulation using ISAJET includes initial and final state parton showers,
fragmentation and decays of heavy flavours, hadronization and finally,
underlying event structure, each of which can potentially impact on the
isolation of leptons from chargino or neutralino decays.
Second, we have taken into account the
breaking of the degeneracy between the squarks and sleptons (and also between
$\tl_{R}$, $\tl_{L}$ and $\tilde{\nu}$) which, as we have mentioned, can have a
significant impact on chargino and neutralino decay patterns\cite{BT,LOPEZ}.

We have used the EHLQ\cite{EHLQ} Set I structure functions in our
computations and have
incorporated the following cuts to simulate the experimental conditions at the
Tevatron.  A toy calorimeter with segmentation
$\Delta\eta\times\Delta\phi = 0.1\times 0.09$
and extending to $|\eta | = 4$ is incorporated. We have assumed an energy
resolution of $70\% /\sqrt{E_T}$ ($15\% /\sqrt{E_T}$) for the
hadronic (electromagnetic)
calorimeter. Jets are defined to be hadron
clusters with $E_T > 15$ GeV in a cone of
$\Delta R = \sqrt{\Delta\eta^2 +\Delta\phi^2} = 0.7$.
Leptons with $E_T > 10$ GeV and within $|\eta | < 3$ are
defined to be isolated\cite{CDF} if there is no hadronic activity exceeding
5 GeV in a cone with $\Delta R = 0.4$ about the lepton direction.
Finally, we require $\eslt > 20$ GeV for all SUSY events.
The events are further classified as follows:
\begin{itemize}
\item ({\it a})
For missing $E_T$ events, we further require $n(jet) \ge 2$, with at least one
jet within $|\eta | < 1$, and $\eslt > 50$ GeV; we veto any event with a jet
within
30 degrees in azimuth of $\vec{\eslt}$, or with an isolated lepton.
If $n(jet) =2$, we require $\eslt > 70$ GeV and veto events with
calorimeter clusters with
$E_T > 5$ GeV within $\pm 30^o$ in azimuth of back-to-back with
the leading jet.
\item ({\it b})
$m$-lepton events are required to contain $m$ isolated leptons in addition to
two or more jets; for $m = 1$ or $2$, we require each lepton to have
$E_T > 15$ GeV and for $m = 2$, we further require
$30^o < \Delta\phi (l_1 ,l_2 ) < 150^o $.
For $m > 2$, we require {\it only} the two fastest leptons to have $E_T > 15$
GeV.
\end{itemize}

The dependence of the various signals on $m_{\tg}$ and $\mu$ is shown
in Fig. 1 and Fig. 2, respectively, for ({\it a}) $m_{\tq} = m_{\tg} + 10$ GeV,
({\it b}) $m_{\tq} = m_{\tg} - 10$ GeV, and ({\it c}) $m_{\tq} = 2m_{\tg}$.
We have fixed $\tan\beta = 2$ and taken $\mu$ ($m_{\tg}$) to
be $-200$ GeV ($200$ GeV) in Fig. 1 (Fig. 2). We have taken $m_{H_p}=500$ GeV
so that all but the lightest Higgs boson are very heavy and
$H_l$ is essentially a SM Higgs boson with a mass in the range
60-80 GeV. We should also note that our calculations of the cross sections
are conservative because, as in any ISAJET simulation, we
specify a limited $p_T$ range for the parent gluinos and squarks.
We have checked that for the heavy
gluino case, the cross sections in Fig. 1 and 2 are essentially unaffected
by this, whereas those for $m_{\tg}$ = 150 GeV case may be underestimated
by as much as 25\%.

We note that the region with positive values of $\mu$ is
essentially excluded\cite{BT} by constraints from experiments at LEP\cite{LEP},
which is why we have confined our attention
to $\mu < 0$ in these figures. We also remark that the CDF
experiment\cite{CDFSUSY} has already excluded gluinos and squarks lighter than
about 90-100 GeV (the squark limit disappears if $m_{\tg} > 400$ GeV), after
cascade decays are incorporated\cite{UPDATE}.

We see that the cross sections in case ({\it c}) are significantly smaller
than in cases ({\it a}) and ({\it b}). This is because
$\tg\tq$ and $\tq\tq$ production processes which dominate
sparticle production in cases ({\it a}) and ({\it b})
are kinematically suppressed for the
case of heavy squarks in ({\it c}).
The difference in magnitudes of the $\eslt$
and $1$-lepton signals of the first two cases is due in part to a 50\% larger
sparticle production cross section for the case ({\it b}) with lighter squarks.
Also, for $m_{\tg}$ = 150 GeV or 200 GeV in Fig. 1b, the sneutrino
mass (as given by the SUGRA relations) is small enough so that
$\tz_2$ ($\tw_1$) dominantly decays via
$\tz_2 \rightarrow \nu\tnu$ ($\tw_1 \rightarrow \l\tnu_l$).

Turning to multilepton production, we see that these cross sections
range up to ${\cal O}(1)$ pb in magnitude for experimentally allowed values of
squark and gluino masses. Thus, a handful of these events may even
be present in current Tevatron experiment data. Since the Tevatron is
expected to continue operation (perhaps even with an increased
luminosity if the main injector is built) for several
years, the observation of these signals with very low SM backgrounds
can lead to a striking confirmation of a SUSY signal in the $\eslt$
channel. To understand the behaviour of the cross sections in Fig. 1,
we first note that for the smaller values of $m_{\tg}$ in case ({\it b}),
dileptons can arise only from the leptonic decays of $\tw_1$ (since,
as we have just seen, $\tz_2$ decays invisibly), while in cases ({\it a})
and ({\it c}) the leptonic decays of the neutralinos (which have enhanced
branching fractions) contribute significantly to the leptonic signals.
This is precisely the reason why the $3$-lepton and $4$-lepton signals
in case ({\it b}) are so small despite the higher production cross section
for gluinos and squarks. The rise in the multilepton cross sections
in Fig. 1b occurs because the invisible $\tz_2\rightarrow \nu\tnu$
decay channel ultimately becomes
kinematically inaccessible with increasing $m_{\tg}$
so that the three-body leptonic decays of the
neutralino become significant. It is also interesting to see that the trilepton
cross section in Fig. 1a exceeds that for same sign dileptons.

We see from Fig. 2 that the cross sections are relatively insensitive to
$\mu$ except in case ({\it b}). The fall in the leptonic cross sections with
increasing values of $|\mu |$ is due to the fact that the $m_{\tw_1}$
and $m_{\tz_2}$ become close to $m_{\tnu}$ so that the leptons tend to
become soft.

The variation of the various signals when $\tan\beta$ is increased from
2 to 20 is illustrated in Table 1 for $m_{\tg} = -\mu = 200$ GeV for
the three cases illustrated in the figures.
There is a general
decrease in the leptonic cross sections for the larger $\tan\beta$ case of
about a factor of 2 in cases ({\it a}) and ({\it c}), except for the trilepton
cross section in case ({\it a}), where the decrease is even greater.
The latter effect is due to a drop in the leptonic branching
fraction of $\tz_2$ for larger values of $\tan\beta$\cite{BT}. For
case (b), the leptonic cross sections are all very small for $\tan\beta = 20$
because $m_{\tw_1}$ and $m_{\tz_2}$ are both only slightly slightly
larger than $m_{\tnu}$, so that $\tz_2$ decays invisibly while the
daughter lepton from the chargino is often too soft to pass the cuts.

We present the $\eslt$ cross section for the cuts discussed above for
comparison with the leptonic signals;
estimates of SM backgrounds to the $\eslt$ signals
may be found in Ref. \cite{CDFSUSY}, along with sparticle mass limit
contours. We see from Fig. 1 that the $1l$ cross-section can be significantly
larger than previous estimates\cite{BTW},
due mainly to the enhanced chargino/neutralino
leptonic branching ratios. The single lepton plus multi-jet signal has large
SM backgrounds from $W+$jets production and $t\bar t$ production. Since the
SUSY $1l$ events are usually accompanied by $3-5$ jets, the best search
strategy
for this channel would be to look for deviations in the $l-\eslt$ transverse
mass distribution of $1l+\ge 4$ jets channel, for which signal can be
comparable to background\cite{GIELE}.
The dominant backgrounds
to the multilepton + jets and the same-sign dilepton signal comes
from the production of top quarks. To estimate these we have generated
20K (10K) t-quark events using ISAJET for $m_t$ = 120 GeV (160 GeV)
and analyzed these using the same cuts as for the signal. From our simulation
we found that just 3 (1) same sign dilepton events, 1 (1) trilepton event
and no 4 lepton events
passed the cuts. Our estimates of the SM backgrounds for these as well as
the dilepton cross sections are shown in Table 2. Combining these results
with those in Fig. 1 and 2, we see that for $m_{\tq} \simeq m_{\tg}$, the SUSY
signal exceeds the background in at least one of the same sign dileptons and
trilepton channels for essentially the whole range of parameters in Fig. 1 and
Fig. 2. Hence, the search for SUSY particles in these channels is only
rate limited.
In particular, given $\sim 1000$ pb$^{-1}$ of data, Tevatron collider
experiments can probe $m_{\tg}\sim 300$ GeV for cases ({\it a})
and ({\it b}).
If squarks are heavy, the same-sign
dileptons remain the most promising of the leptonic signals. It is
worth emphasizing that the leptonic signals involving muons may be
significantly larger than shown in the figures
since detection of muons with $p_T \leq 10$ GeV
should be possible.

In summary,we have shown that the production of squarks and gluinos
can lead to an observable rate for spectacular event topologies, including
same sign dileptons, $3l$ and $4l$ + $jets$ + $\eslt$ events at the
Fermilab Tevatron. The rates are largest if squarks and gluinos are
approximately degenerate as is the case in many models\cite{SPECTRA}.
We have also estimated SM backgrounds to these event topologies and
found them to be small. Signals from 100-300 GeV gluinos and squarks
should certainly be detectable
at the Tevatron, particularly if the main injector upgrade is approved.
For the favourable case, where the squark is just heavier than the gluino,
a handful of such events could even be present in the current data
sample. Unambiguous detection of these gluino and squark signals would
also provide evidence
for charginos and neutralinos at the Tevatron collider.

%
\acknowledgements
We thank Sharon Hagopian for discussions, and Lupe Howell for technical
support.
This research was supported in part by the U.~S. Department of Energy under
contract number DE-FG05-87ER40319 and DE-AM03-76SF00235.
\bigskip

{\it Note added:} After finishing this manuscript, we received a related
paper on same-sign dilepton signals from gluino pair production\cite{NEWG}.
\newpage
%
%
%
%

%
\newpage
%
%
\figure{Total cross sections for $\eslt$, $1l$, $2l$, $3l$, $4l$ and
same-sign (SS) dilepton event topologies
in $p\bar p$ collisions at $\sqrt s=1.8$~TeV, for cuts discussed in
the text
versus $m_{\tg}$. We have fixed $\mu =-200$ GeV, $\tan\beta =2$, $m_t =140$
GeV,
$m_{H_p}=500$ GeV and ({\it a}) $m_{\tq}=m_{\tg}+10$ GeV,
({\it b}) $m_{\tq}=m_{\tg}-10$ GeV,
and ({\it c}) $m_{\tq}=2m_{\tg}$. The slepton masses are determined in terms
of $m_{\tq}$ and $m_{\tg}$ using renormalization group equations to
evolve from a common sfermion mass at the unification scale. The
error bars reflect the statistical uncertainty in our simulation of
10K events for each point. All values of $m_{\tg}$ shown are consistent
with LEP data except in case ({\it b}) where $m_{\tnu} \geq 40$ GeV requires
$m_{\tg} \geq 165$ GeV. The straight line segments between the data points are
to guide the eye.
\label{FIG1}}
%
\figure {The dependence of the cross sections for various event
topologies on $\mu$ for $m_{\tg}=200$ GeV for the same cuts and
model parameters as in Fig. 1. LEP data require $\mu \leq -80$ GeV.
\label{FIG2}}
\eject
%
%
\begin{table}
\caption{Cross sections in pb at Tevatron after cuts for $\tan\beta =2$ and 20.
Other parameters are as in Fig. 1. We did not find any multilepton events
for $m_{\tq}=190$ GeV, $\tan\beta =20$ case. The bound signifies the 1 event
level in our simulation of 5K events for each $\tan\beta =20$ point.}
\begin{tabular}{cccccccc}
$m_{\tg}$ & $m_{\tq}$ & $\tan\beta$ & $\eslt$ & 1-l & 2-l
& 3-l & SS \\
\tableline
200 & 210 & 2  & 3.3 & 2.2 & $7.8\times 10^{-2}$ &
0.12 & $4.7\times 10^{-2}$ \\

200 & 210 & 20 & 3.7 & 1.9 & $8.0\times 10^{-2}$ &
$1.0\times 10^{-2}$ & $3.3\times 10^{-2}$ \\

200 & 190 & 2  & 6.5 & 1.3 & $2.3\times 10^{-2}$ &
$3.8\times 10^{-3}$ & $5.7\times 10^{-3}$ \\

200 & 190 & 20 & 7.0 & $7.6\times 10^{-3}$ & $<3.8\times 10^{-3}$ &
$<3.8\times 10^{-3}$ & $<3.8\times 10^{-3}$ \\

200 & 400 & 2  & 0.60 & 0.26 & $1.9\times 10^{-2}$ &
$1.1\times 10^{-3}$ & $5.9\times 10^{-3}$ \\

200 & 400 & 20 & 0.57 & 0.18 & $8.3\times 10^{-3}$ &
$5.5\times 10^{-4}$ & $2.2\times 10^{-3}$ \\

\end{tabular}
\label{TABLE1}
\end{table}
%
%
%
\begin{table}
\caption{Multi-lepton background in pb from $t\bar t$ production
(after cuts) }
\begin{tabular}{ccccc}
$m_t$ & 2-l & 3-l & 4-l & SS \\
\tableline
120 & $0.38\pm 0.02$ & $(1\pm 1)\times 10^{-3}$ & $<1\times 10^{-3}$ &
$(3.9\pm 2.2)\times 10^{-3} $ \\
160 & $0.14\pm 0.01$ & $ (6\pm 6)\times 10^{-4}$ & $<6\times 10^{-4}$ &
$(6\pm 6)\times 10^{-4}$ \\
\end{tabular}
\label{TABLE2}
\end{table}
%
%
%
\end{narrowtext}
\end{document}